\documentclass[a4paper,12pt]{article}

\usepackage[left=2.5cm,right=2.5cm,top=2.5cm,bottom=2.5cm]{geometry}
\usepackage{pslatex}
\usepackage{textpos}
\usepackage{caption}
\usepackage{listings}
\usepackage{color}
\usepackage{textcomp}

\usepackage{epsfig}

\usepackage{pgf}
\usepackage{xcolor}
\usepackage{amsmath,amssymb,amsfonts,amsthm,breqn,amsbsy}
\usepackage{longtable}
\usepackage{verbatim}
\usepackage{float}
\usepackage[title]{appendix}
\usepackage{array}
\usepackage{varwidth}
\usepackage{multirow}
\usepackage{tabularx}
\usepackage{framed}
\usepackage{lipsum}
\usepackage[round]{natbib}
\usepackage[breaklinks = true, hidelinks]{hyperref}
\usepackage{breakcites}
\usepackage{longtable}
\usepackage{booktabs}
\usepackage{lscape}
\usepackage{rotating}
\usepackage{fancyvrb}
\usepackage{setspace}
\usepackage[T1]{fontenc}
\usepackage{indentfirst}
\usepackage{lineno}
\usepackage{mathtools}
\usepackage{bigints}
\usepackage{xurl}
\usepackage{bm,bbm}
\usepackage[normalem]{ulem}

\makeatletter
\newcommand*\mysize{%
  \@setfontsize\mysize{9.0}{9.0}%
}
\makeatother

\title{Exchangeable Gaussian Processes with application to epidemics}
\author{Lampros Bouranis$^{\ast,1}$, Petros Barmpounakis$^2$, Nikolaos Demiris$^1$, \\ Konstantinos Kalogeropoulos$^3$}

\makeatletter
\renewcommand\@date{{%
  \large\centering
  $^{1}$\textit{Department of Statistics, Athens University of Economics and Business, Athens, Greece}
  \par
  \vspace{0.5ex}
  $^{2}$\textit{Department of Oncology, University of Cambridge, Cambridge, United Kingdom}
  \par
  \vspace{0.5ex}
  $^{3}$\textit{Department of Statistics, The London School of Economics and Political Science, London, United Kingdom}
  \par
  \vspace{0.5ex}
  $^{\ast}$bouranis@aueb.gr
}}
\makeatother

\begin{document}
\maketitle

\begin{abstract}
We develop a Bayesian nonparametric framework for modeling collections of related epidemic trajectories using exchangeable Gaussian processes. The proposed formulation induces structured dependence across time-varying transmission processes through shared latent Gaussian process components, leading to interpretable pooling behaviour, variance decompositions and intra-class correlation parameters that quantify information sharing across populations. The framework accommodates both direct and indirect observation regimes, including noisy measurements of effective reproduction numbers, latent transmission models for Chikungunya outbreaks and age-structured epidemic systems driven by partially observed infection dynamics. We discuss connections with broader multi-output Gaussian process formulations, including coregionalisation and latent factor models, while emphasising the interpretation of the proposed approach as a functional random-effects framework for related temporal trajectories. A simulation study investigates the behaviour of the proposed models under varying cross-population dependence structures including dependent, weakly dependent and independent settings. Across several epidemic applications, the proposed models demonstrate strong predictive performance while providing interpretable insights into shared and population-specific transmission dynamics.
\end{abstract}

\noindent%

\section{Introduction}\label{section:intro}
Epidemic datasets often contain several temporal trajectories corresponding to different countries, regions, islands or age groups. Although such trajectories may exhibit substantial heterogeneity, they are also expected to share some common features, reflecting biological, behavioural and environmental mechanisms that drive disease transmission. Statistical models for such datasets should therefore allow for flexible time-varying behaviour while simultaneously facilitating borrowing of information across related trajectories. Gaussian process (GP) priors provide a natural framework
for modelling latent time-varying processes and their
associated uncertainty. In this paper, we focus on
the development of structured multi-output GP models
for epidemic trajectories, motivated by the
hierarchical and exchangeable structures often arising
in infectious disease applications. 

Multi-task (or `multi-output') Gaussian process models have received considerable attention in machine learning and statistics; see \cite{alvarez} for a review. A large part of the literature has focused on models for panels of related temporal or spatial trajectories, often constructed through latent processes combined linearly across tasks. Examples include the coregionalisation models \citep{Journel,Goovaerts} in geostatistics and the semi-parametric latent factor models of \cite{teh}. In many applications, however, the structure of the data and scientific considerations may motivate more structured forms of dependence between the trajectories. Examples include formulations based on shared latent mean processes for multi-task forecasting \citep{leroy} and Gaussian process models for epidemic transmission dynamics \citep{dureau,kypraios,ghosh,bouranis}. Such formulations may be particularly appealing in settings where the observed processes correspond to related populations and where the main inferential interest lies in modelling shared and population-specific temporal variation.

Our methodological contribution in this paper is concerned
with the development of exchangeable Gaussian process models
for panels of epidemic trajectories. The proposed formulation
is motivated by settings where the trajectories correspond
to different but related populations and each trajectory
may be viewed as arising from an underlying population
process. The resulting hierarchical formulation facilitates
borrowing of information across trajectories while allowing
for population-specific temporal dynamics. The proposed
models may also be viewed as structured special cases within
the broader class of coregionalisation and latent factor
models, with the imposed structure leading to parsimonious
and interpretable dependence formulations. In particular,
the proposed framework enables decomposition of variability
into shared and population-specific components and
facilitates learning about the degree of shrinkage and
dependence across trajectories.

In epidemic applications, the observed data may consist
of noisy, partial, and indirect measurements of the
underlying transmission dynamics. Different data streams
may be linked to latent transmission-related processes
through distinct observation mechanisms, including
nonlinear stochastic epidemic models. We therefore
distinguish the epidemiological models describing
transmission and observation processes from the
exchangeable GP priors used to specify the joint
temporal dependence of their latent components.
The resulting framework accommodates both Gaussian
and nonlinear observation models and is applicable
more generally to longitudinal and other temporally
indexed data.

Fitting models of epidemic trajectories may range from relatively straightforward settings involving directly observed temporal processes to substantially more challenging problems where the transmission dynamics are only partially observed through nonlinear stochastic epidemic systems. We exploit the structure of the proposed formulations to derive efficient computational schemes. In particular, we consider empirical Bayes inference
based on marginal likelihood optimisation for
directly observed epidemic trajectories, and fully
Bayesian inference using MCMC for nonlinear epidemic
models with indirectly observed latent transmission
processes. We further discuss computational and scalability aspects of the proposed framework, including formulations based on Markov kernels, state-space representations and Kalman filtering methods, as well as limitations arising in more general covariance structures and settings involving covariates. The proposed methodology is assessed through simulation studies under varying dependence structures and through applications to epidemic datasets related to the Chikungunya arbovirus and SARS-CoV-2.

The remainder of the paper is organised as follows. Section \ref{section:epi_models} introduces the epidemic models and data motivating the proposed methodology. 
Section \ref{section:methods} presents the exchangeable GP formulations and discusses their connections with structured multi-output GPs. 
Section \ref{section:inference} develops the associated inference procedures and discusses computational and scalability aspects of the proposed framework.
Section \ref{section:simulation} presents simulation studies under different dependence structures, while Section \ref{section:applications} contains applications to epidemic datasets. We conclude the paper in Section \ref{section:discussion} with some discussion.


\section{Epidemic data and observation models}
\label{section:epi_models}

Epidemic datasets often contain collections of related temporal trajectories corresponding to different countries, regions, islands or age groups. Although such trajectories may exhibit substantial heterogeneity, they are also expected to share common features reflecting biological, behavioural and environmental mechanisms that drive disease transmission. Statistical models for such settings should therefore balance borrowing of information across related populations with sufficient flexibility to capture genuine between-population heterogeneity. This work is motivated by outbreak datasets related to the Chikungunya arbovirus and the severe acute respiratory syndrome coronavirus 2 (SARS-CoV-2), which naturally contain grouped temporal structures. In particular, we consider settings where the observed trajectories correspond to related populations and where the underlying transmission dynamics may reasonably be assumed to share common temporal characteristics while retaining population-specific deviations.

A central modelling challenge in each application is determining the appropriate degree of pooling across populations. Fitting independent temporal models ignores common epidemic dynamics driven by the same pathogen, overlapping interventions or shared environmental pressures, and may yield unstable estimates when surveillance data are sparse or noisy. Conversely, complete pooling may mask important structural differences such as heterogeneous national policies and population immunities in the cross-country COVID-19 setting, island-specific environmental effects in the Chikungunya outbreaks, or age-dependent transmission, contact and mortality patterns in age-stratified epidemic analyses. Partial pooling provides a natural compromise by allowing each population-specific trajectory to deviate from a shared latent process, with the degree of sharing learned from the data.

In the applications considered below, the latent transmission mechanisms are represented through time-varying stochastic processes. The use of Gaussian processes and related diffusion formulations for modelling epidemic transmission dynamics is now well established in the epidemic modelling literature, see for example \cite{dureau,kypraios,ghosh,bouranis}. In this work, we focus on introducing structured exchangeable GP formulations for collections of related epidemic trajectories.

\subsection{COVID-19 reproduction numbers in Europe}
\label{section:logRt_model}

A widely used measure for tracking epidemic dynamics is the effective reproduction number $R_t^{eff}$ \citep{diekmann,davies,knock}. 
The quantity $R^{eff}_{i,t}$ can be thought of as the average number of secondary infections generated by an infected individual at day $t$ in country $i$, accounting for population immunity and depletion of susceptibles. It is typically considered as a threshold parameter since taking values above or below unity suggest a growing or declining epidemic respectively. In this application, a task corresponds to a country-specific trajectory of the log-transformed effective reproduction number. We consider posterior median estimates of the daily effective reproduction number for  Greece, the United Kingdom, Germany, Czech Republic, Denmark, Estonia, Finland, France, Ireland, Poland and Spain, obtained from the framework of \cite{abbott1,abbott2}. Let $y_{i,t}$ denote the observed log-transformed reproduction number for country $i$ at day $t$. 
We assume
\begin{equation}
\label{eq:rt_observation}
y_{i,t}\mid x_{i,t},\sigma_y
\sim N(x_{i,t},\sigma_y^2),
\end{equation}
where $x_{i,t}$ denotes a latent reproduction-number trajectory and $\sigma_{y} \in \mathbb{R}^{+}$ is an unknown parameter. 
In Section \ref{section:methods}, the latent trajectories $\{x_{i,t}\}$ are assigned exchangeable Gaussian process priors in order to model dependence and pooling across countries.

\subsection{Chikungunya epidemics in French Polynesia and the French West Indies}
\label{section:chikv_model}

\cite{riou_chikv} investigated the drivers of transmission
and variability in epidemic waves caused by Chikungunya
and Zika viruses using data from territories in French
Polynesia and the French West Indies. Their framework
combines stochastic transmission dynamics with
meteorological covariates affecting disease spread.
In this work, we focus on the Chikungunya outbreaks
and adopt a time-dependent susceptible--infectious--recovered
formulation linking transmission and observation processes. Let $O_{s,t}$ denote the observed incidence count on
island $s$ for week $t$, and let $\beta_{s,t}$ denote
the corresponding island-specific transmission rate.
The baseline model of \cite{riou_chikv} allows for
an island-specific random effect $b_s$ in the
transmission term, together with lagged precipitation
effects extending up to eight weeks:
\begin{equation}
\begin{cases}
\log\beta_{s,t}
=
b_s+\displaystyle\sum_{l=0}^{8}
P_{s,t-l}\log\beta_{P,s,l},
\\[2mm]
b_s\sim\mathcal{N}(\mu_B,\sigma_B^2),
\end{cases}
\label{eq:chikv_baseline}
\end{equation}
where $P_{s,t-l}$ denotes precipitation at lag $l$
and $\beta_{P,s,l}$ is the corresponding coefficient.

We extend this formulation by replacing the
time-invariant random effects with latent
time-varying transmission processes:
\begin{equation}
\log\beta_{s,t}
=
x_{s,t}
+
\sum_{l=0}^{8}
P_{s,t-l}\log\beta_{P,s,l},
\label{eq:chikv_transmission}
\end{equation}
where $x_{s,t}$ represents the latent time-varying
transmission trajectory on the log scale.
The joint temporal dependence of these latent
processes across islands is specified through
the exchangeable GP framework introduced in
Section \ref{section:methods}.

Given the transmission rate, the expected number
of new infections on island $s$ at time $t$ is
\begin{equation}
d_{s,t}
=
\beta_{s,t}O^*_{s,t}
\left(
1-\frac{\sum_{\tau=1}^{t-1}O_{s,\tau}}{N_s}
\right),
\label{eq:chikv_expected}
\end{equation}
where $O^*_{s,t}$ denotes the exposure process
and $N_s$ is the population size of island $s$. The observed incidence counts $O_{s,t}$ are then
modeled using a Negative Binomial distribution,
allowing for overdispersion in the surveillance data:
\begin{equation}
O_{s,t}\mid\theta
\sim
\operatorname{NegBin}
(d_{s,t},\xi_{O,s,t}),
\label{eq:chikv_observation}
\end{equation}
where $\xi_{O,s,t}=d_{s,t}/\phi_O$, with
$\phi_O>0$ denoting the overdispersion parameter.
Under this parameterisation,
$\operatorname{Var}(O_{s,t}\mid\theta)
=d_{s,t}(1+\phi_O)$.

An important epidemiological quantity derived from
the model is the island-specific effective reproduction
number, which characterises the evolution of
transmission over time. It is given by
\begin{equation}
R_s^{\mathrm{eff}}(t)
=
z_{s,t}\exp\{x_{s,t}\},
\label{eq:chikv_reproduction}
\end{equation}
where
\[
z_{s,t}
=
1-\frac{\sum_{\tau=1}^{t-1}O_{s,\tau}}{N_s}
\]
is the proportion of the population of island $s$
remaining susceptible to Chikungunya infection
at time $t$.

\subsection{Age-stratified COVID-19 data in England}
\label{section:multiseir}
\sloppy{
Our third application considers age-stratified COVID-19 mortality data from England \citep{uk_datasets}. Following \cite{flaxman,monod,chatzilena,bouranis}, we separate the epidemic system into a latent transmission process and an observation process in order to account for noisy daily surveillance data and delayed reporting mechanisms. The population is partitioned into $A$ age groups such that $\mathbb{N} = \sum_{\alpha = 1}^{A}\mathbb{N}_{\alpha}$. The available information also contains a contact matrix $C$ of dimension $A \times A$ whose element $C_{\alpha,\alpha'}$ represents the average number of contacts between individuals of age group $\alpha$ and age group $\alpha'$. The contact matrix is scaled by the average infectious period, $v_{I}$, which is assumed to be the same across all age groups during the time horizon of the analysis.
}

We begin by specifying the transmission process. The transmissibility is related to a latent process through
\begin{equation}
\beta_{\alpha,t}
=
\operatorname{logit}^{-1}(x_{\alpha,t}),
\label{eq:covid_transmissibility}
\end{equation}
where $x_{\alpha,t}$ represents the latent
age-specific transmissibility trajectory on the
logit scale. The model based counts of new daily age-specific infections are then given by
\begin{equation}
\label{eq:covid_infections}
i_{\alpha, t} = s_{\alpha, t}\beta_{\alpha, t} \sum_{\alpha' = 1}^{A}
C_{\alpha,\alpha'}v_{I}
\bigg(
\sum_{\tau = 1}^{t-1}
i_{\alpha', \tau} g_{t - \tau}
\bigg),
\end{equation}
where $s_{\alpha, t} = 1 - \frac{\sum_{\tau=1}^{t-1} i_{\alpha,\tau} }{\mathbb{N}_{\alpha}}$ is the proportion of the population in age band $\alpha$ that remains susceptible to SARS-CoV-2 infection. 
The generation time follows a Gamma distribution with mean 6.5 days and coefficient of variation 0.62, $g^{CT} \sim \text{Gamma}(6.5, 0.62)$ discretised to days: $g(t) = \int^{t+0.5}_{t-0.5}g^{CT}(u)du, \forall~2,3,\ldots$ and $g(1) = \int^{1.5}_{0}g^{CT}(u)du$ for $t = 1$.

To complete the model, we specify the observation process linking latent infections to observed mortality. Let $y_{\alpha, t}$ be the number of observed deaths on day $t = 1, \ldots, T$ in age group $\alpha \in \{1,\ldots,A\}$, which is linked
with the expected number of new infections according to the equation
\begin{equation}
\label{eq:covid_expected_deaths}
d_{\alpha, t} = \mathbb{E}[y_{\alpha, t}] = \text{IFR}_{\alpha} \times \sum_{s = 1}^{t-1}i_{\alpha, s} h_{t-s} 
\end{equation}
where $d_{\alpha, t}$ are the new expected age-stratified mortality counts, $\text{IFR}_{\alpha}$ denotes the infection fatality rate for age group $\alpha$ and $h$ the infection-to-death distribution. Here $h_{s}$ denotes the probability of death $s$ days after infection; $h$ is assumed to be Gamma distributed with mean 24.2 days and coefficient of variation 0.39 \citep{flaxman}. The age-specific infection fatality ratio estimates are informed by large-scale prevalence surveys.

Finally, the observed mortality counts are modelled using a Negative Binomial distribution to account for overdispersion \citep{riou1,birrell,seaman}:
\begin{equation}
y_{\alpha,t}\mid\theta
\sim
\operatorname{NegBin}
(d_{\alpha,t},\xi_{D,\alpha,t}),
\label{eq:covid_observation}
\end{equation}
where $\xi_{D,\alpha, t} = \frac{d_{\alpha, t}}{\phi_D}$ such that $\mathbb{V}[y_{\alpha, t}] = d_{\alpha, t}(1+\phi_D)$ with $\phi_D \in \mathbb{R}^{+}$ the overdispersion parameter.

\section{Exchangeable Gaussian process models}
\label{section:methods}

The epidemiological models introduced in Section \ref{section:epi_models} share a common feature: each involves a collection of latent time-varying processes associated with
related populations. Specifically, these are
\begin{equation*}
\begin{aligned}
x_{i,t} &: \quad
\text{Reproduction number trajectory for country } i,
\\
x_{s,t} &: \quad
\text{log-transmission trajectory for island } s,
\\
x_{\alpha,t} &: \quad
\text{logit-transmissibility trajectory for age group } \alpha.
\end{aligned}
\end{equation*}
Although their epidemiological interpretations and
observation models differ, all three applications
require a specification of the joint temporal
dependence of these latent processes across
populations. To provide a common framework, we consider
$p$ related latent processes
$\{x_i(t)\}_{i=1}^{p}$.
For the zero-mean specification used in our
applications, we write
\begin{equation}
\mathbf{x}(\cdot)\mid\theta
\sim
\mathcal{GP}\bigl(\mathbf{0},K_\theta\bigr),
\label{eq:multioutput_gp}
\end{equation}
where $\mathbf{x}(t)=(x_1(t),\ldots,x_p(t))^\top$
and $K_\theta(t,t')$ is a matrix-valued covariance
function describing dependence across populations
and time. We obtain this joint GP specification
through a hierarchical exchangeable construction,
in which each latent trajectory is decomposed into
a shared temporal component and a population-specific
deviation.

\subsection{Model construction}
\label{section:models}

We construct a class of exchangeable Gaussian process models for collections of related temporal trajectories. The proposed formulation is motivated by settings where the latent processes correspond to different but related populations and are assumed to share common temporal characteristics while retaining population-specific deviations. The baseline version of the model can be constructed in two steps:

\begin{enumerate}
    \item Assume there exists an underlying latent stochastic process $\mu(t)$, with $t\in(0,T]$, representing characteristics shared across a collection of stochastic processes denoted by $x_{i}(t), i={1,\dots,p}$. We define a Bayesian non-parametric model for $\mu(t)$ using a GP prior.

    \item Define the stochastic process $D_{i}(t)$ as the population-specific deviation between $x_{i}(t)$ and $\mu(t)$, i.e.
    $$
    D_{i}(t):=x_{i}(t)-\mu(t), \;\forall i,
    $$
    with each $D_i(t)$ also modelled using GP priors.
\end{enumerate}

We assume that $\mu(t)$ follows a GP with mean function $m_{\mu}(t;\beta)$ and covariance kernel of the form
\begin{equation}
\mu(t)\sim \mathcal{GP}\left(m_{\mu}(t;\beta),
\sigma_{\mu}^2 k_{\mu}(t,t')\right),
\label{eq:kernelform}
\end{equation}
where $m_{\mu}(t;\beta)$ may correspond to a deterministic or parametric mean specification, $\sigma_{\mu}^2 \in \mathbb{R}^{+}$ is an unknown scale parameter and $k_{\mu}(t,t^{\prime})$ denotes a covariance kernel. For simplicity, the applications considered in this work use zero mean functions unless otherwise stated. Examples of covariance kernels include the squared exponential kernel
$$
k_{\mu}(t,t^{\prime}) = \exp\left(-\frac{(t-t^{\prime})^2}{\ell}\right),\; \ell \in (0, \infty),
$$
and the Brownian motion kernel
$$
k_{\mu}(t,t^{\prime})=\min (t,t^{\prime}).
$$

The shared process $\mu(t)$ can be integrated out analytically, leading to a marginal representation which provides additional insight into the induced dependence structure and facilitates inference. This can be done in two different ways depending on whether observations are available at common equidistant time points across populations.

\begin{itemize}

\item \textbf{Case of full observation at equidistant points:}
We begin by assuming (and later relax this assumption) that we have equidistant observations at times $t_1,\dots,t_n=T$ and denote the vectors consisting of the values of $\mu(t)$ and $x_{i}(t)$ at those times with
$$
M=(\mu(t_1),\dots,\mu(T))^{\prime}
$$
and
$$
X_i=(x_{i}(t_1),\dots,x_{i}(T))^{\prime},
$$
respectively. We write
$$
X_i \mid M \sim N\big(M,\sigma_{x}^2C_{x}\big),\;\;\;\forall i=1,\dots,p,
$$

\begin{equation}
M\sim N\big(0_n,\sigma_{\mu}^2C_{\mu}\big),
\label{eq:Mmarginal}
\end{equation}
where $C_{x}$ and $C_{\mu}$ correspond to the covariance matrices generated by applying the GP kernels for $x_{i}(t)$ and $\mu(t)$ respectively, assuming common parameter $\sigma_{x}^2 \in \mathbb{R}^{+}$ for each $i$ without loss of generality.

Next, consider the vector $X$ obtained by stacking all the $X_i$'s one after the other. We write
$$
X \mid M \sim N\big(1_p\otimes M, \;\sigma_{x}^2 I_p\otimes C_{x}\big),
$$
where $\otimes$ denotes the Kronecker product, $1_p$ is a vector of $p$ ones and $I_p$ denotes the identity matrix of dimension $p$. Standard properties of the multivariate normal distribution then provide the marginal distribution of $X$ as
\begin{equation}
X \sim N\big( 0_{np}, \;\sigma_{\mu}^2 1_p 1_p^{\prime}\otimes C_{\mu} +\sigma_{x}^2 I_p\otimes C_{x}\big).
\label{eq:Amarginal}
\end{equation}

\item \textbf{General case:}
The assumption of equidistant observations can be removed if we work directly with covariance kernels. Each function $x_i(t)$ may be written as the sum
$$
x_i(t) = D_i(t) + \mu(t),
$$
where $\mu(t)$ is a shared GP
$$
\mu(t) \sim \mathcal{GP}(0, \sigma_\mu^2 k_\mu(t,t')),
$$
and each $D_i(t)$ is an independent draw from
$$
\mathcal{GP}(0, \sigma_x^2 k_x(t,t')).
$$
Under this construction the resulting marginal GP has zero mean function and covariance kernel given by
\begin{equation}
\label{eq:general_kernel}
k_{i,j}(t,t') = E[ x_i(t) x_j (t')] 
= \delta_{i,j} \sigma_x^2 k_x(t,t') + \sigma_\mu^2 k_\mu(t,t'),
\end{equation}
where $i,j=1,\dots,p$, $\delta_{i,j}$ denotes
the Kronecker delta, and
$K_\theta(t,t')=[k_{i,j}(t,t')]_{i,j=1}^{p}$
is the matrix-valued covariance function
introduced above. It is not hard to verify that the kernel in \eqref{eq:general_kernel} leads to \eqref{eq:Amarginal} in the case of equidistant observations, while also accommodating more general observation schemes.

\end{itemize}

\subsection{Intra-class correlation, shrinkage and interpretability}

The covariance structure in \eqref{eq:general_kernel} induces a natural decomposition of variability into shared and population-specific components. In particular, $\sigma_\mu^2$ controls the variability associated with the shared latent process $\mu(t)$, while $\sigma_x^2$ controls the variability associated with the population-specific deviations $D_i(t)$. This decomposition provides an interpretation analogous to variance components in hierarchical and multi-level models, but defined at the level of stochastic processes.

The proposed framework also induces a natural parameter governing the strength of cross-sectional dependence between trajectories. In particular, when the kernels satisfy
$$
k_\mu(t,t)=k_x(t,t)=1,
$$
as is the case for squared exponential kernels written in correlation form, the cross-sectional correlation is given by
\begin{equation}
\rho
=
\frac{\sigma_\mu^2}
{\sigma_\mu^2+\sigma_x^2}.
\label{eq:icc}
\end{equation}
The same expression is obtained under common Brownian motion scaling, since in that case
$$
k_\mu(t,t)=k_x(t,t)=t.
$$
The parameter $\rho$ therefore plays a role analogous to the intra-class correlation coefficient encountered in hierarchical and multi-level models and determines the degree of pooling induced across trajectories. Values of $\rho$ close to one correspond to settings where the trajectories are dominated by the shared latent process. In such cases, the model induces substantial pooling of information across populations and trajectories shrink strongly towards the common latent structure. 
In contrast, values of $\rho$ close to zero correspond
to weakly related or nearly independent trajectories,
with inference driven primarily by the
population-specific processes. Importantly,
exchangeability does not imply positive dependence:
the independent GP model is a special case of the
proposed exchangeable construction when
$\sigma_\mu^2=0$ (equivalently, $\rho=0$).
Thus, $\rho$ quantifies the strength of
cross-population dependence rather than the
degree of exchangeability. More generally, the
temporal dependence structure also depends on the
covariance kernels $k_\mu(t,t')$ and $k_x(t,t')$,
which determine the smoothness and persistence
properties of the shared and population-specific
components respectively. Nevertheless, the
parameter $\rho$ continues to provide a useful
interpretable summary of the relative contribution
of shared versus individual-specific variability
and of the overall degree of cross-population
dependence.

The proposed formulation therefore retains features commonly encountered in hierarchical and random-effects models while operating directly on stochastic processes. More general formulations may allow the population-specific components to possess different variance and covariance parameters across populations. In such settings, the framework naturally induces population-specific shrinkage and dependence parameters through the corresponding variance decompositions. In epidemic applications this is particularly relevant when assessing the extent to which transmission trajectories across countries, regions or age groups exhibit common temporal behaviour, thus benefiting from pooling of information across populations. Consequently, such inference may lead to suitably differentiated tailored policies and/or interventions across these groups.

\subsection{Connection with coregionalisation and latent factor models}

The proposed formulation is closely related to the broader class of multi-output GP models based on coregionalisation and latent factor constructions \citep{bonilla2008multi,alvarez}. In particular, the covariance structure in \eqref{eq:general_kernel} is a structured
special case of the linear model of
coregionalisation (LMC), commonly used in multi-task Gaussian process modelling. Specifically, the shared process induces a rank-one cross-population covariance component, while the population-specific deviations contribute a diagonal component, corresponding to a restricted coregionalisation structure. Under the proposed decomposition
$$
x_i(t)=\mu(t)+D_i(t),
$$
the shared latent process $\mu(t)$ induces dependence across trajectories, while the processes $D_i(t)$ capture population-specific temporal variation. The resulting covariance structure therefore leads naturally to interpretable variance decompositions and pooling parameters associated with shared and population-specific variability.

To provide further intuition, consider the analogy with standard random-effects models. A random-intercept model induces a covariance structure of the form
\[
\operatorname{Var}(y_i)
=
\sigma_b^2 1_{n_i}1_{n_i}^\top
+
\sigma^2 I_{n_i},
\]
which may be viewed as a restricted latent factor covariance representation. Nevertheless, random-effects models are widely used as statistical formulations in their own right, owing to their hierarchical interpretation, parsimony, and inferential focus. Similarly, the proposed framework may be viewed as a functional random-effects formulation for collections of related temporal trajectories, emphasising structured pooling, variance decomposition, and interpretable dependence across populations, rather than highly flexible latent factor representations. In epidemic applications, this is particularly relevant when assessing the extent to which transmission dynamics are shared across related populations.

\subsection{Multi-level extensions}

The proposed framework can be extended to incorporate multiple hierarchical layers. In epidemic applications such layers may arise from data collected across countries, regions within countries, age groups within regions, or other nested population structures. The corresponding exchangeable GP model may then be constructed recursively by assigning GP priors to successive levels of population-specific deviations. For example, one may begin with a common latent process representing global transmission, followed by country-specific and then region- or age-specific deviation processes. In this work we focus primarily on two-level formulations in order to study the implications of exchangeability, pooling and interpretable dependence structures in epidemic trajectories. Extensions to more general hierarchical settings are nevertheless conceptually straightforward within the proposed framework.

\section{Inference, computation and scalability}\label{section:inference}

\subsection{Observation models and posterior structure}

The exchangeable GP formulations introduced in Section \ref{section:methods} define the latent stochastic process component of the model. To complete the specification, the latent trajectories are coupled with an observation model for the data and prior distributions for the remaining static parameters. Let $Y$ denote the observed data, $X$ the collection of latent trajectories and $\theta$ the vector of static parameters. The resulting posterior distribution may be written generically as
\begin{equation}
\label{eq:general_posterior}
\pi(X,\theta \mid Y)\propto f(Y \mid X,\theta)\pi(X \mid \theta)\pi(\theta),
\end{equation}
where $f(\cdot)$ denotes the observation model(s) introduced in Section \ref{section:epi_models}. Depending on the application, the latent trajectories $X$ may either correspond directly to observed quantities measured with noise, or may represent latent transmission processes embedded within nonlinear stochastic epidemic systems. The proposed framework accommodates both Gaussian and non-Gaussian observation structures. In Gaussian settings, substantial analytical simplifications become available through marginalisation of latent GP components, leading to efficient likelihood-based inference schemes. In non-linear epidemic models however, the latent transmission dynamics are analytically intractable and inference relies on Monte Carlo-based Bayesian procedures. The following subsections discuss these two inferential regimes in turn.

\subsection{Marginalisation and Gaussian inference}

The marginalisation of latent GP components is particularly relevant in applications where the trajectories are observed directly, possibly with measurement noise. Inference on the shared latent process $\mu(t)$ can be carried out through the conditional distribution
$$
\pi(M \mid X,\theta,Y)=\pi(M \mid X,\theta),
$$
which follows from the conditional independence structure of the model. In particular,
\begin{equation}
\label{eq:Mposterior}
M \mid X,\theta \;\sim \; N\bigg((\Sigma^{-1}_{\mu} + p\Sigma^{-1}_{x})^{-1}\Sigma^{-1}_{x}\sum^{p}_{i=1}X_i,  \;\;(\Sigma^{-1}_{\mu} + p\Sigma^{-1}_{x})^{-1}\bigg),
\end{equation}
where $\Sigma_\mu$ and $\Sigma_x$ denote the covariance matrices associated with the shared and population-specific GP components respectively. 

Additional simplifications arise in the presence of Gaussian observation models. For example, and without loss of generality, suppose that
\begin{equation*}
Y\sim N(X,\sigma_y^2 I_n),
\end{equation*}
where $Y$ is the vector obtained by concatenating the observations across populations. Marginalising over the latent trajectories then yields
\begin{equation}
\label{eq:YMarginal}
Y\mid \theta \;\sim\; N(0,K + \sigma_y^2 I_n),
\end{equation}
where $K=\text{Cov}(X)$. The resulting Gaussian structure leads to closed-form posterior and predictive distributions. In particular,
\begin{equation}
\label{eq:APosterior}
X\mid Y, \theta \;\sim\; N(K(K + \sigma_y^2 I_n)^{-1}Y ,K- K(K + \sigma_y^2 I_n)^{-1}K),
\end{equation}
while the predictive distribution for future observations $Y^*$ is given by
\begin{equation}
\label{eq:Predictive}
Y^*\mid Y,\theta \;\sim\; N\bigg(K_{Y^*Y}(K + \sigma_y^2 I_n)^{-1}Y, K_{Y^*Y^*}- K_{Y^*Y}(K + \sigma_y^2 I_n)^{-1}K_{Y^*Y}\bigg),
\end{equation}
where $K_{Y^*Y^*}$ denotes the covariance matrix of $Y^*$ and $K_{Y^*Y}$ the cross-covariance matrix between $Y$ and $Y^*$. See Section~S1 of the Supplementary Material of more details.

The marginal likelihood formulation in \eqref{eq:YMarginal} enables efficient likelihood-based inference schemes based either on Bayesian posterior simulation or on direct optimisation of the marginal likelihood with respect to the covariance parameters, corresponding to Type II maximum likelihood estimation. The latter approach is used extensively in the simulation studies and Gaussian applications considered in this work.

\subsection{Bayesian inference for nonlinear epidemic models}

In nonlinear epidemic models, the latent transmission dynamics are typically only partially observed and analytical marginalisation of the latent processes is not generally available. In such settings, inference must be based on simulation-based Bayesian procedures targeting the posterior distribution in \eqref{eq:general_posterior}. The proposed framework is compatible with a range of inferential methodologies for partially observed stochastic systems. In this work, we focus primarily on Markov chain Monte Carlo (MCMC) schemes applied to marginal formulations of the model whenever partial analytical simplifications are available. In particular, we consider Hamiltonian Monte Carlo (HMC) methods implemented through the Stan software framework \citep{homan,carpenter2017stan}, which are well suited to high-dimensional hierarchical models involving strongly correlated latent structures.

\subsection{Computation, Markov kernels and scalability}

The computational requirements of the proposed framework depend on the temporal covariance kernels, the number of populations and observations, and the observation model. Although generic dense GP calculations for $p$ trajectories on $T$ time points can require $O((pT)^3)$ operations, the exchangeable covariance structure permits substantial simplifications in balanced settings. For trajectories observed on a common grid, the covariance matrix is
\[
K=\mathbf{1}_p\mathbf{1}_p^\top\otimes K_\mu+I_p\otimes K_D,
\]
where $K_\mu$ and $K_D$ are the temporal covariance matrices of the shared and population-specific components. The exchangeable structure allows the covariance matrix to be block-diagonalised into one $T\times T$ matrix, $pK_\mu+K_D$, and $p-1$ identical blocks, $K_D$. Thus, for balanced Gaussian observation models with compatible noise structures, the required covariance matrix factorisations involve matrices of dimension $T$ rather than $pT$, reducing their computational cost from $O((pT)^3)$ to $O(T^3)$.

Further computational advantages arise for temporal kernels admitting finite-dimensional Markov representations, reflecting the broader connection between Gaussian processes, stochastic differential equations (SDE) and sparse precision structures \citep{lindgren2011,sarkka2019}. The Brownian motion specification used in our simulations and epidemic applications has an explicit exchangeable SDE representation. Writing $X_i(t)=\mu(t)+D_i(t)$, with independent Brownian motions $W_0,W_1,\ldots,W_p$, gives
\[
d\mu(t)=\sigma_\mu\,dW_0(t),\qquad
 dD_i(t)=\sigma_D\,dW_i(t),
\]
so that
\[
dX_i(t)=\sigma_\mu\,dW_0(t)+\sigma_D\,dW_i(t),\qquad
\operatorname{Cov}\{dX(t)\}=\bigl(\sigma_\mu^2\mathbf{1}_p\mathbf{1}_p^\top+\sigma_D^2 I_p\bigr)\,dt.
\]
This representation allows the exchangeable latent processes to be treated within a linear Gaussian state-space framework, enabling efficient Kalman filtering and smoothing under Gaussian observation models. A stationary extension is obtained by replacing the Brownian motion components with independent Ornstein--Uhlenbeck processes,
\[
\begin{aligned}
d\mu_t &=-\lambda_\mu(\mu_t-m)\,dt+\sqrt{2\lambda_\mu}\sigma_\mu\,dW_{0,t},\\
dD_{i,t} &=-\lambda_D D_{i,t}\,dt+\sqrt{2\lambda_D}\sigma_D\,dW_{i,t}.
\end{aligned}
\]
With stationary initial distributions, the resulting exchangeable process has covariance
\[
\operatorname{Cov}(X_i(t),X_j(s))=\sigma_\mu^2 e^{-\lambda_\mu|t-s|}+\mathbb{I}(i=j)\sigma_D^2 e^{-\lambda_D|t-s|}.
\]
When $\lambda_\mu=\lambda_D=\lambda$, the vector $\mathbf X_t$ itself satisfies
\[
d\mathbf X_t =-\lambda(\mathbf X_t-m\mathbf 1_p)\,dt+\sqrt{2\lambda}\,\Sigma^{1/2}d\mathbf B_t,
\]
\sloppy{
where $\Sigma=\sigma_\mu^2\mathbf 1_p\mathbf 1_p^\top+\sigma_D^2 I_p$. For different mean-reversion rates, the augmented state $(\mu_t,D_{1,t},\ldots,D_{p,t})^\top$ remains Markovian and admits an exact linear Gaussian state-space representation. The corresponding transition distributions, including their derivation, are provided in Section~S2 of the Supplementary Material.
}

The computational implications differ for nonlinear or non-Gaussian epidemic observation models. In these settings, the latent Markov structure remains available, but standard Gaussian Kalman filtering is not directly applicable to the full observation model. Particle filtering, particle MCMC and other state-space inference methods can exploit this structure, providing alternatives to computations involving dense temporal covariance matrices. State-space and stochastic-process formulations have been used in partially observed epidemic models with a range of inference methods, including particle-based \citep{dureau}, approximation-based \citep{ghosh} and Hamiltonian Monte Carlo \citep{bouranis} approaches. Non-Markovian kernels, including the squared-exponential (RBF) kernel, do not generally admit exact finite-dimensional state-space representations. Nevertheless, finite-dimensional SDE approximations can be constructed for such kernels, including the RBF kernel \citep{hartikainen2010}. On regular temporal grids, stationary covariance functions also induce Toeplitz matrices, allowing circulant-embedding and FFT-based methods to accelerate Gaussian simulation and covariance operations. Related structured computations have been used in Hamiltonian MCMC for partially observed SDEs driven by fractional Brownian motion \citep{beskos2015}. Such approaches provide alternatives to direct dense covariance computations.

For the Gaussian observation models in the simulation study and the European $R_t$ application, we use Type II maximum likelihood implemented in GPflow, with fitting times typically of a few minutes for the datasets considered. The nonlinear epidemic applications use data-augmented MCMC implemented in RStan, see Supplementary Material Section~S3 for details. These implementations demonstrate the computational feasibility of the proposed framework at the scales considered in this paper. More generally, the computational strategies discussed above apply to a broad range of Gaussian process models used in epidemic modelling, including the covariance structures and temporal grids considered in our applications. More generally, the state-space and structured-covariance methods discussed above apply to a broad range of Gaussian process models used in epidemic modelling, including the covariance structures and temporal grids considered in our applications. These properties provide opportunities for scalable inference without necessarily introducing approximations to the underlying covariance models. The choice of temporal kernel also raises questions about the trade-off between scalability and predictive performance, which we investigate through the simulations in Section~5 and the empirical applications in Section~6. In particular, Brownian motion provides the best predictive performance in Section~6.3 and performs comparably to the RBF kernel in Section~6.2, while the RBF kernel performs better in Section~6.1. For settings where these computational structures are unavailable, further gains may be achieved through approximate GP methods, including induced-point schemes \citep{snelson,titsias}, low-rank and sparse representations and variational methods \citep{quinonero,Liu2018}.

\section{Simulation study}\label{section:simulation}

The simulation study investigates the predictive performance of the proposed exchangeable Gaussian process (xGP) framework under varying degrees of cross-population dependence and model misspecification. The primary objective is to explore its behaviour across a range of dependence structures, addressing the practical question of when it is beneficial to borrow strength across related stochastic processes. Since the prequential design requires repeated model fitting at successive forecasting origins, inference is performed using the computationally efficient empirical Bayes (Type II maximum likelihood) implementation.

\subsection{Simulation design}

We considered four data-generating mechanisms representing varying degrees of cross-dependence and covariance heterogeneity. Throughout, we simulated panels consisting of $p=10$ stochastic processes observed over $T=100$ equally spaced time points. The first $50$ observations were used as an initial training period, after which predictions were derived sequentially every seven time points. Forecasts were evaluated at horizons $h=1$ and $h=7$, with the latter representing the primary forecasting task considered in the main text; results for $h=7$ are reported in Supplementary Material Section~S4.

In all scenarios the observation model was
\[
y_i(t)=x_i(t)+\varepsilon_i(t),
\qquad
\varepsilon_i(t)\sim N(0,\sigma_y^2),
\]
with common observation variance across all models and simulation settings. The latent processes were generated according to one of the following mechanisms.

\begin{enumerate}

\item Substantial cross-dependence:
The latent trajectories followed the exchangeable decomposition
\[
x_i(t)=\mu(t)+D_i(t),
\]
%
%
\[
\mu(\cdot)\sim GP(0,\sigma_\mu^2K_{\ell_\mu}),
\qquad
D_i(\cdot)\sim GP(0,\sigma_x^2K_{\ell_x}),
\]
with $\sigma_\mu^2=2$ and $\sigma_x^2=1$, corresponding to an intra-class correlation
of $2/3$.

\item Weak cross-dependence: The same hierarchical model was used, but with
\[
\sigma_\mu^2=0.1,
\qquad
\sigma_x^2=1,
\]
producing only weak dependence between trajectories (ICC approximately equal to $0.09$).

\item Independent trajectories: The latent processes were generated independently using GPs with common covariance parameters across trajectories.

\item Heterogeneous independent trajectories: The latent processes were generated independently with trajectory-specific covariance parameters, representing a setting with neither cross-dependence nor having a common covariance structure across trajectories.

\end{enumerate}

\subsection{Models compared}

Four models were fitted to every simulated dataset:

\begin{itemize}

\item xRBF: exchangeable GP with common kernel parameters,

\item mxRBF: exchangeable GP with trajectory-specific kernel parameters,

\item iRBF: independent GPs with common kernel parameters,

\item miRBF: independent GPs with trajectory-specific kernel parameters.

\end{itemize}

Model fitting was performed using Type II maximum likelihood (empirical Bayes), optimising the marginal likelihood with respect to the covariance hyperparameters. This choice reflects the intended forecasting application of the methodology and permits efficient repeated model fitting within the prequential evaluation framework.

\subsection{Evaluation}

Predictive performance was assessed using a prequential forecasting framework. Following the initial training period, the models were re-estimated sequentially as new observations accrue. At each forecasting origin, predictive distributions were obtained for the specified forecasting horizon. The primary performance criterion was the cumulative joint logarithmic score, obtained by summing the log predictive densities across all forecasting origins within each Monte Carlo replication. This provides an overall measure of predictive evidence accumulated during the sequential forecasting exercise, analogous to a cumulative log predictive density. For each simulation scenario, the cumulative log scores were averaged across the 50 Monte Carlo replications. Monte Carlo standard errors and corresponding 95\% confidence intervals were computed from the empirical variability across replications. The Mean Absolute Error (MAE) and the Root mean squared error (RMSE) were also reported as secondary measures of forecasting accuracy.

\subsection{Results}

Table~\ref{tab:simulation} summarises the predictive performance of the competing models.

\begin{table}[H]
\centering
\caption{Prequential forecasting performance in the simulation study for the RBF models at the one-week-ahead horizon. Values are Monte Carlo means across 50 simulated datasets, with 95\% Monte Carlo confidence intervals in parentheses. Lower values indicate better predictive performance.}
\label{tab:sim_prequential_rbf_h7}
\resizebox{\textwidth}{!}{
\begin{tabular}{llccc}
\hline
Scenario & Model & Joint Log Score & MAE & RMSE \\
\hline

Substantial cross-dependence
& xRBF & \textbf{1071.3 (1064.1, 1078.5)} &\textbf{ 0.956 (0.944, 0.967)} & \textbf{1.200 (1.186, 1.214)} \\
& mxRBF & 1093.5 (1085.0, 1102.0) & 0.981 (0.968, 0.994) & 1.232 (1.216, 1.247) \\
& iRBF & 1092.8 (1084.5, 1101.2) & 0.983 (0.970, 0.996) & 1.235 (1.219, 1.251) \\
& miRBF & 1116.6 (1105.7, 1127.5) & 1.010 (0.995, 1.025) & 1.269 (1.250, 1.287) \\
\hline

Weak cross-dependence
& xRBF & \textbf{1066.5 (1059.6, 1073.3)} & \textbf{0.948 (0.937, 0.958)} & \textbf{1.184 (1.171, 1.196)} \\
& mxRBF & 1086.5 (1077.4, 1095.6) & 0.968 (0.956, 0.980) & 1.211 (1.196, 1.226) \\
& iRBF & 1067.2 (1060.3, 1074.1) & 0.948 (0.938, 0.959) & 1.185 (1.172, 1.197) \\
& miRBF & 1085.0 (1076.1, 1093.8) & 0.966 (0.954, 0.977) & 1.209 (1.194, 1.223) \\
\hline

Independent, common kernel
& xRBF & 1108.6 (1102.0, 1115.3) & 1.008 (0.998, 1.017) & 1.264 (1.252, 1.277) \\
& mxRBF & 1137.7 (1129.8, 1145.5) & 1.043 (1.032, 1.053) & 1.309 (1.294, 1.323) \\
& iRBF & \textbf{1108.4 (1101.8, 1115.1)} &\textbf{ 1.007 (0.997, 1.017)} & \textbf{1.264 (1.252, 1.276)} \\
& miRBF & 1136.6 (1128.6, 1144.6) & 1.041 (1.030, 1.053) & 1.307 (1.293, 1.322) \\
\hline

Independent, heterogeneous kernels
& xRBF & 1048.9 (1042.1, 1055.8) & 0.915 (0.907, 0.924) & 1.152 (1.140, 1.163) \\
& mxRBF& 1045.8 (1038.2, 1053.5) & 0.911 (0.902, 0.920) & 1.144 (1.132, 1.157) \\
& iRBF & 1049.4 (1042.4, 1056.4) & 0.915 (0.907, 0.924) & 1.152 (1.141, 1.164) \\
& miRBF & \textbf{1043.0 (1035.8, 1050.1)} & \textbf{0.907 (0.898, 0.916)} & \textbf{1.139 (1.128, 1.151)} \\
\hline
\end{tabular}
}
\label{tab:simulation}
\end{table}

A number of consistent patterns emerge. First, in the presence of substantial cross-dependence (Scenario~1), the exchangeable GP achieves the best predictive performance, demonstrating the benefit of borrowing strength across related trajectories. Second, under weak cross-dependence (Scenario~2), exchangeable and independent models exhibit very similar performance, indicating that the proposed model remains robust even when only limited shared structure is present. Third, when the data are generated independently but with common covariance parameters (Scenario~3), the exchangeable model incurs essentially no loss in predictive performance, demonstrating that imposing the exchangeable formulation has little predictive cost when the data provide little evidence of shared structure. Finally, under heterogeneous independent trajectories (Scenario~4), independent models achieve slightly better predictive performance, as expected under genuine model misspecification. Nevertheless, the performance degradation of the exchangeable model remains modest. Overall, the simulation study indicates that exchangeable GPs provide substantial gains whenever meaningful shared structure exists while remaining remarkably robust when cross-dependence assumptions are only weakly satisfied or moderately violated. 

To further investigate the behaviour of the empirical Bayes estimation procedure, we summarised the fitted intra-class correlation coefficient (ICC) obtained from the final model fit in each Monte Carlo replication for the two cross-dependence simulation scenarios. Table~\ref{tab:icc_recovery} reports the true ICC together with summaries of the empirical distribution of the fitted ICC estimates across the 50 Monte Carlo replications. Broadly speaking, the fitted ICC estimates exhibit reasonable agreement with the corresponding true values despite the complexity of the hierarchical GP model and the well-known challenges associated with estimating variance components. In particular, the fitted ICCs clearly distinguish between the substantial and weak cross-dependence settings, suggesting that the empirical Bayes estimation procedure is able to recover the underlying degree of cross-dependence with satisfactory accuracy.

\begin{table}[ht]
\centering
\caption{Empirical distribution of the fitted ICC estimates across the 50 Monte Carlo replications for the two cross-dependence simulation scenarios.}
\label{tab:icc_recovery}
\begin{tabular}{lccccc}
\hline
Scenario & True ICC & Mean & 2.5\% & Median & 97.5\%\\
\hline
Substantial cross-dependence & 0.667 & 0.573 & 0.313 & 0.607 & 0.824\\
Weak cross-dependence   & 0.091 & 0.104 & 0.000 & 0.080 & 0.289\\
\hline
\end{tabular}
\end{table}

We also repeated the simulation study using Brownian motion kernels (Supplementary Material, Section S4). The resulting performance rankings were qualitatively identical to those obtained under the RBF kernel, indicating that the conclusions are not sensitive to the particular covariance kernel adopted.

\section{Real-data Applications}\label{section:applications}

The proposed methodology is illustrated using three
epidemic applications that highlight complementary
aspects of exchangeable GPs across different dependence
structures and observation models. The first considers
sequential prediction of reproduction numbers across
European countries, demonstrating the predictive benefits of exchangeable formulations and the interpretability of heterogeneous, country-specific pooling parameters. The remaining two applications embed exchangeable GPs within nonlinear epidemic models and employ full Bayesian inference. The Chikungunya application illustrates how the framework can characterise relatively weak dependence across island-specific transmission processes, despite
modest differences in leave-one-out predictive performance between exchangeable and independent models. In contrast, the analysis of age-stratified COVID-19 mortality in England reveals substantial shared temporal variation across age groups, accompanied by substantial improvements in leave-one-out predictive performance relative to independent formulations. Together, these applications complement the simulation study by illustrating how the extent of cross-population dependence varies across epidemiological settings and how the proposed functional random-effects construction provides interpretable information about shared and population-specific transmission dynamics.

Throughout this section, iBM/xBM denote independent and
exchangeable Gaussian process models with Brownian motion
kernels, respectively, while iRBF/xRBF denote the
corresponding models with squared exponential kernels.
Prefixing a model by $m$ (e.g.\ miRBF or mxRBF)
indicates that kernel parameters are allowed to
differ across populations. Full implementation
details for all three applications are provided
in Supplementary Section~S3.


\subsection{Prequential analysis - COVID-19 reproduction number in Europe}
\label{section:prequential}

We first consider a prequential forecasting study based on daily posterior median estimates of the effective reproduction number, $R_t$, obtained by \cite{abbott2,abbott1} for eleven European countries: Greece, the United Kingdom, Germany, Czech Republic, Denmark, Estonia, Finland, France, Ireland, Poland and Spain. The logarithm of the reported $R_t$ estimates was modelled using the GP formulations introduced in Section~\ref{section:logRt_model}. Specifically, we use the Gaussian observation model in \eqref{eq:rt_observation}, with the latent trajectories assigned exchangeable or independent GP priors as described in Section~\ref{section:models}. This application provides a real-data counterpart to the simulation study and allows us to investigate whether the predictive benefits of exchangeable GPs observed under controlled conditions also arise in an out-of-sample forecasting exercise. The first 63 observations, from 10 December 2021 to 11 February 2022, were used for initial model fitting. We then adopted a prequential evaluation framework \citep{dawid}: models were re-estimated every seven days using all observations available at each forecasting origin and used to generate one-week-ahead forecasts over a six-week evaluation period ending on 25 March 2022.

\begin{table}[ht]
\centering
\caption{Prequential forecasting performance for the European $R_t$ data. Lower values indicate better predictive performance.}
\label{tab:prequential_results}
\begin{tabular}{lccc}
\toprule
Model & Joint Log Score & MAE & RMSE\\
\midrule
mxRBF & \textbf{-1862.0} & \textbf{0.0067} & \textbf{0.0138}\\
xRBF  & -1621.9 & 0.0081 & 0.0155\\
miRBF & -1809.9 & 0.0072 & 0.0147\\
iRBF  & -1583.9 & 0.0081 & 0.0163\\
\midrule
mxBM & -385.5 & 0.0584 & 0.1060\\
xBM  & -151.0 & 0.0577 & 0.1085\\
miBM & -327.2 & 0.0581 & 0.1066\\
iBM  & -136.1 & 0.0575 & 0.1087\\
\bottomrule
\end{tabular}
\end{table}

Table~\ref{tab:prequential_results} shows that the multiple exchangeable model equipped with an RBF kernel achieves the best predictive performance according to the cumulative logarithmic score. The same model also attains the smallest MAE and RMSE, indicating improved point prediction in addition to superior probabilistic forecasting. Overall, the RBF models substantially outperform their Brownian motion counterparts, indicating that the smoother temporal dependence induced by the RBF kernel is better suited to these reproduction-number trajectories.

The results are also consistent with the main findings of the simulation study. There, exchangeable formulations improved prediction when dependence across trajectories was present, including settings in which the degree of dependence was relatively weak. The present analysis provides a complementary empirical example: allowing information to be shared across country-specific trajectories improves both probabilistic and point prediction relative to the corresponding independent RBF formulations. 
\begin{figure}[H]
\centering
\includegraphics[width=0.92\textwidth]{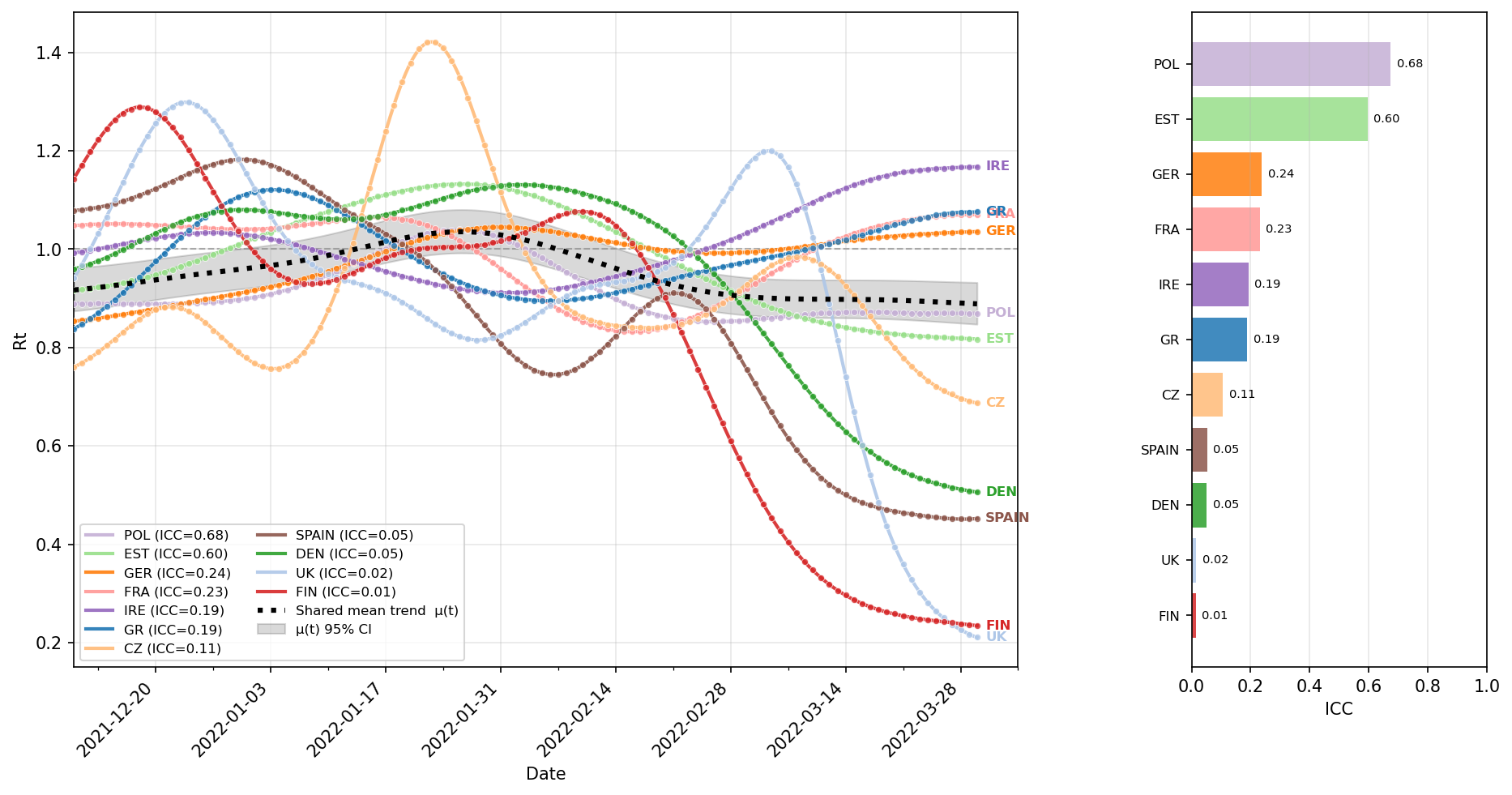}
\caption{
Estimated trajectories under the preferred multiple exchangeable RBF model for the European reproduction number data.
The coloured curves correspond to the fitted country-specific latent trajectories, while the black dashed line denotes the estimated common European trajectory. The common trajectory captures the temporal component shared across countries, whereas the country-specific trajectories represent local departures from this shared epidemic evolution.
}
\label{fig:rt_trajectories}
\end{figure}
Figure~\ref{fig:rt_trajectories} presents the fitted country-specific trajectories together with the estimated common European trajectory under the multiple-exchangeable RBF model. The latter represents the latent temporal component shared across countries under the exchangeable model. For example, Poland, which has the largest estimated ICC ($\rho=0.68$), remains relatively close to the common component over much of the study period. In contrast, Finland has the smallest estimated ICC ($\rho=0.01$) and exhibits a pronounced country-specific departure towards the end of the study, with its reproduction number declining substantially below the common trajectory. More generally, the estimated ICCs vary substantially across countries, from $0.68$ for Poland and $0.60$ for Estonia to $0.02$ for the United Kingdom and $0.01$ for Finland, suggesting considerable heterogeneity in the extent to which country-specific dynamics conform to the shared European component. This heterogeneity also provides an interpretation for the improved forecasting performance of the multiple-exchangeable RBF model relative to its counterpart with common kernel parameters.

\subsection{Chikungunya epidemics in French Polynesia and the French West Indies}
\label{section:chikv_analysis}

We analyse weekly incidence data from Chikungunya outbreaks occurring between 2013 and 2016 in territories of French Polynesia and the French West Indies \citep{riou_chikv}. The datasets consist of surveillance counts collected through local sentinel networks of health practitioners and naturally define collections of related epidemic trajectories across islands and archipelagoes. Daily precipitation measurements were aggregated to weekly measurements and incorporated as covariates in the transmission process in order to account for potential environmental effects on disease spread. We use the transmission and incidence models specified in equations~\eqref{eq:chikv_transmission}--\eqref{eq:chikv_observation} of Section~\ref{section:chikv_model}, assigning either exchangeable GP models with a shared latent
component or independent GP models to the latent transmission trajectories as described in Section~\ref{section:models}.

The aim of this application is to investigate the extent to which latent transmission dynamics are shared across geographically related outbreaks, while allowing for island-specific temporal variation. Unlike the preceding application, transmission is not observed directly but is inferred from incidence data through a nonlinear epidemic model. The exchangeable GP construction provides an interpretable decomposition of shared and island-specific variation, allowing us to assess the strength of dependence across islands even when differences in leave-one-out predictive performance between competing formulations are modest.

Table~\ref{tab:chikv_loo_table} compares the baseline transmission model with the GP formulations. All GP models achieve substantially lower LOOIC values than the baseline, highlighting the importance of time-varying transmission. Differences among the GP models are relatively small, with no clear advantage for the exchangeable formulations. The xBM model has the lowest LOOIC for French Polynesia, whereas mxRBF performs best for the French West Indies.

\begin{figure}[H]
\centering
\includegraphics[scale=0.5]{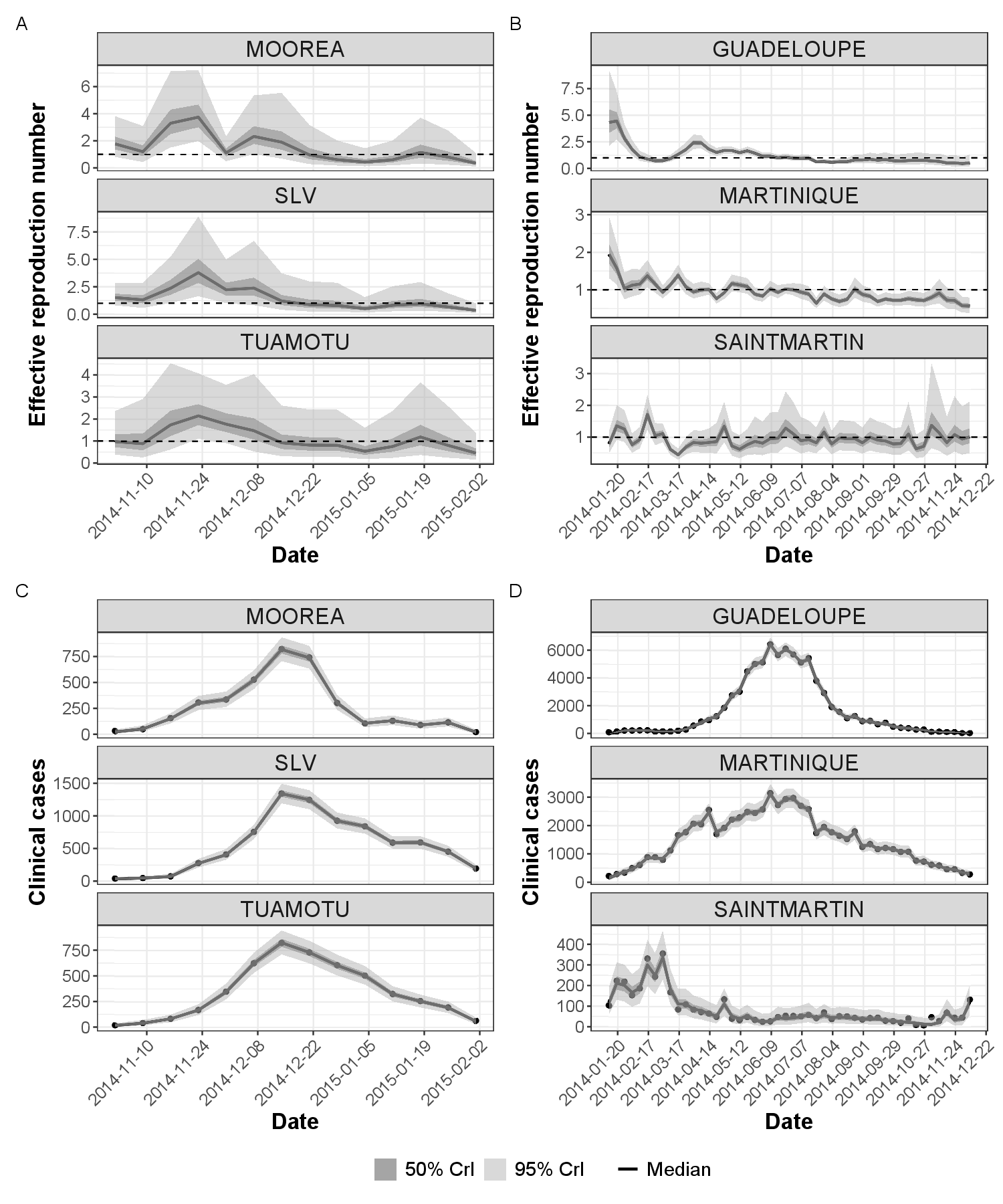}
\caption{Chikungunya epidemics in French Polynesia (Panels A and C) and French West Indies (Panels B and D). 
Estimates for French West Indies are based on the mxRBF transmission model, while estimates for French Polynesia are based on the xBM transmission model. Panels A and B: island-specific effective reproduction number. Panels C and D: goodness-of-fit against the weekly number of incident cases reported (points). Model estimates are based on posterior medians, together with 50\% and 95\% credible intervals (CrIs) of draws from the posterior predictive distribution.}
\label{fig:CHIKV_FP_FWI_2015_mainTextPlots}
\end{figure}

Figure~\ref{fig:CHIKV_FP_FWI_2015_mainTextPlots} presents the estimated island-specific effective reproduction numbers, $R_s^{eff}(t)$, together with their posterior uncertainty and the corresponding fits to the observed incidence data. In French Polynesia, the estimated reproduction numbers exceed
one during the early stages of the outbreaks, consistent with the sharp increase in infections observed in December 2014. In the French West Indies, transmissibility generally declines in Guadeloupe and Martinique, while Saint Martin exhibits more persistent fluctuations around the epidemic threshold. Panels C and D show the fitted incidence trajectories together with their pointwise posterior credible intervals, indicating reasonable agreement with the observed data. Together with the ICC estimates, these results illustrate how the exchangeable GP framework can quantify shared transmission dynamics while retaining island-specific variation and uncertainty.
Table~\ref{tab:chikv_icc} reports the posterior distributions of the intra-class correlation coefficients (ICCs) for the two island groups, quantifying the amount of cross-population dependence. Under the xBM model, the posterior mean ICC is 0.510 for French Polynesia, compared with 0.078 for the French West Indies.
These estimates suggest some degree of cross-dependence in French
Polynesia, albeit with considerable posterior uncertainty, and
much weaker cross-dependence in the French West Indies. The
multiple-exchangeability models reveal further heterogeneity in
the degree of sharing across islands, particularly in French
Polynesia. Thus, even when differences in model fit are small,
the exchangeable formulation provides useful information about
the dependence structure of the latent transmission processes.
\begin{table}[H]
\caption{Chikungunya epidemics -- Model predictive performance.
$\Delta\text{LOOIC}$: expected LOOIC difference between each model and the model with the lowest $\text{LOOIC}$ across the model set.
All estimates are accompanied by their respective standard errors where applicable.
}
\centering
\setlength{\tabcolsep}{0.20em}
\begin{tabular}{llrrr}
\toprule
\textbf{Location} 
& \textbf{Model} 
& \textbf{LOOIC (se)}
& \textbf{$\Delta\text{LOOIC}$ (se)} 
& \textbf{CPU (sec)} \\
\hline			
Polynesia 
& Baseline 
& 1028.3 (33.4)
& 281.9 (34.7) 
& 47.42 \\
& miBM  
& 748.8 (15.9) 
& 2.4  (3.8)   
& 403.05   \\
& xBM  
& 746.4 (14.8) 
& \textit{Ref} 
& 1013.79  \\
& mxBM 
& 753.8 (15.1) 
& 7.4  (4.6)   
& 1010.38  \\
& miRBF   
& 755.6 (17.1) 
& 9.2  (6.7)   
& 553.71   \\
& xRBF 
& 753.1 (15.0) 
& 6.8  (4.4)  
& 612.33   \\
& mxRBF  
& 753.3 (14.2) 
& 7.0  (3.7)   
& 969.69   \\
\hline	
West Indies  
& Baseline  
& 2158.1 (102.2) 
& 423.0 (96.6) 
& 58.53   \\
& miBM      
& 1747.8 (69.6) 
& 12.6 (10.3)  
& 680.02  \\
& xBM      
& 1757.7 (62.3)  
& 22.5 (14.6)  
& 540.35  \\
& mxBM     
& 1747.1 (71.1)  
& 11.9 (12.0)  
& 346.39  \\
& miRBF       
& 1763.9 (65.8) 
& 28.7 (12.7)  
& 1360.71 \\
& xRBF      
& 1757.8 (58.3)  
& 22.6 (13.0)  
& 1093.79 \\
& mxRBF      
& 1735.2 (65.9) 
& \textit{Ref} 
& 2533.96 \\
\bottomrule
\end{tabular}
\label{tab:chikv_loo_table}
\end{table}

\begin{table}[htbp]
\caption{Posterior means and 95\% credible intervals for the
intra-class correlation coefficients (ICCs) in the Chikungunya
application. FP: French Polynesia; FWI: French West Indies.}
\centering
\small
\begin{tabular}{llcc}
\toprule
\textbf{Model} & \textbf{ICC / Island}
& \textbf{FP} & \textbf{FWI} \\
\midrule
xBM & Overall
& 0.510 (0.094, 0.843) & 0.078 (0.000, 0.294) \\
\midrule
mxBM
& Tahiti / Saint-Martin
& 0.644 (0.028, 0.999) & 0.085 (0.000, 0.254) \\
& Moorea / Martinique
& 0.227 (0.007, 0.643) & 0.359 (0.001, 0.982) \\
& SLV / Guadeloupe
& 0.300 (0.009, 0.786) & 0.078 (0.000, 0.252) \\
& Tuamotu / --
& 0.662 (0.033, 0.999) & -- \\
& Marquises / --
& 0.452 (0.050, 0.956) & -- \\
\midrule
xRBF & Overall
& 0.543 (0.027, 0.917) & 0.085 (0.020, 0.230) \\
\midrule
mxRBF
& Tahiti / Saint-Martin
& 0.613 (0.010, 0.989) & 0.088 (0.024, 0.227) \\
& Moorea / Martinique
& 0.282 (0.003, 0.770) & 0.192 (0.051, 0.505) \\
& SLV / Guadeloupe
& 0.369 (0.004, 0.883) & 0.042 (0.009, 0.128) \\
& Tuamotu / --
& 0.622 (0.010, 0.997) & -- \\
& Marquises / --
& 0.514 (0.029, 0.988) & -- \\
\bottomrule
\end{tabular}
\label{tab:chikv_icc}
\end{table}


\subsection{Age-stratified COVID-19 deaths in England}
\label{section:covid19_analysis}

We discuss findings for the period March 2, 2020 to September 27, 2020, before vaccine introduction. We use the age-stratified epidemic model specified in equations~\eqref{eq:covid_transmissibility}--\eqref{eq:covid_observation} of Section~\ref{section:multiseir}, either exchangeable GP models with a shared latent
component or independent GP models to the latent
transmission trajectories as described in Section~\ref{section:models}. We analyse reported deaths for England, divided into four age groups, $\{0-19, 20-39, 40-59, 60+\}$, accounting for the age distribution of the general population, $\mathbb{N}$.
\begin{table}[H]
\caption{COVID-19 epidemic in England -- Model predictive performance.
$\Delta\text{LOOIC}$: expected LOOIC difference between each model and the model with the lowest $\text{LOOIC}$ across the model set.
All estimates are accompanied by their respective standard errors where applicable.
}
\centering
\begin{tabular}{llrrr}
\toprule
\textbf{Model} 
& \textbf{LOOIC (se)}
& \textbf{$\Delta\text{LOOIC}$ (se)}  
& \textbf{CPU (hrs)}\\
\hline								
miBM  
& 3367.4 (107.4)  
& 488.0 (45.7) 
& 18.37 \\
xBM  
& 2879.3 (95.3) 
& \textit{Ref}   
& 27.47 \\
mxBM 
& 2882.8 (95.5)  
& 3.5 (3.9)    
& 24.01 \\
miRBF   
& 3017.6 (101.6) 
& 138.3 (155.6) 
& 27.93\\
xRBF   
& 3026.6 (101.2) 
& 147.3 (155.4) 
& 38.33 \\
mxRBF  
& 3028.2 (101.4) 
& 148.9 (155.5) 
& 34.24 \\
\bottomrule
\end{tabular}
\label{tab:covid19_seir_loo_table}
\end{table}
We adopted the country-specific synthetic contact matrix of \cite{prem1}, constructed based on national demographic and socioeconomic characteristics (Supplementary Material, Section S5).
In the absence of a synthetic contact matrix for England we used the one created for the United Kingdom, a plausible assumption due to large overlap. The elements of the contact matrix represent the daily average contact rates between age groups.
Total cases were estimated via the age-stratified infection fatality rate (IFR), informed by the REACT-2 national prevalence study of over 100,000 adults in England \citep{ward_react2}.
\begin{table}[H]
\caption{Posterior means and 95\% credible intervals for the
intra-class correlation coefficients (ICCs) in the
age-stratified COVID-19 application in England.}
\centering
\small
\begin{tabular}{llc}
\toprule
\textbf{Model} & \textbf{Parameter / Age group}
& \textbf{ICC (95\% CrI)} \\
\midrule
xBM & Overall & 0.922 (0.858, 0.964) \\
\midrule
mxBM
& 0--19 & 0.825 (0.664, 0.928) \\
& 20--39 & 0.990 (0.935, 1.000) \\
& 40--59 & 0.994 (0.965, 1.000) \\
& 60+ & 0.945 (0.846, 0.990) \\
\midrule
xRBF & Overall & 0.674 (0.463, 0.847) \\
\midrule
mxRBF
& 0--19 & 0.794 (0.587, 0.898) \\
& 20--39 & 0.921 (0.903, 1.000) \\
& 40--59 & 0.989 (0.958, 1.000) \\
& 60+ & 0.935 (0.827, 0.983) \\
\bottomrule
\end{tabular}
\label{tab:covid19_icc}
\end{table}
Table~\ref{tab:covid19_seir_loo_table} shows that the exchangeable BM models achieve substantially lower LOOIC values than the independent BM model, with xBM attaining the lowest LOOIC. The difference between xBM and mxBM is small relative to its standard error, suggesting little additional benefit from allowing age-specific pooling parameters. The RBF models perform less well overall, with no clear advantage for exchangeable over independent formulations within this kernel family. The posterior ICC estimates in Table~\ref{tab:covid19_icc} provide further evidence of strong cross-dependence across age groups. Under the xBM model, the posterior mean ICC is 0.922 (95\% credible interval: 0.858--0.964), indicating that most of the latent temporal variation in transmission is shared. The mxBM estimates suggest particularly strong sharing among the adult age groups, with ICCs ranging from 0.945 to 0.994, while the youngest group (0--19 years) has a somewhat lower ICC of 0.825. This pattern suggests greater similarity in the temporal transmission dynamics of adults than between adults and children, although the posterior uncertainty should be taken into account.

\begin{figure}[H]
\centering
\includegraphics[scale=0.42]{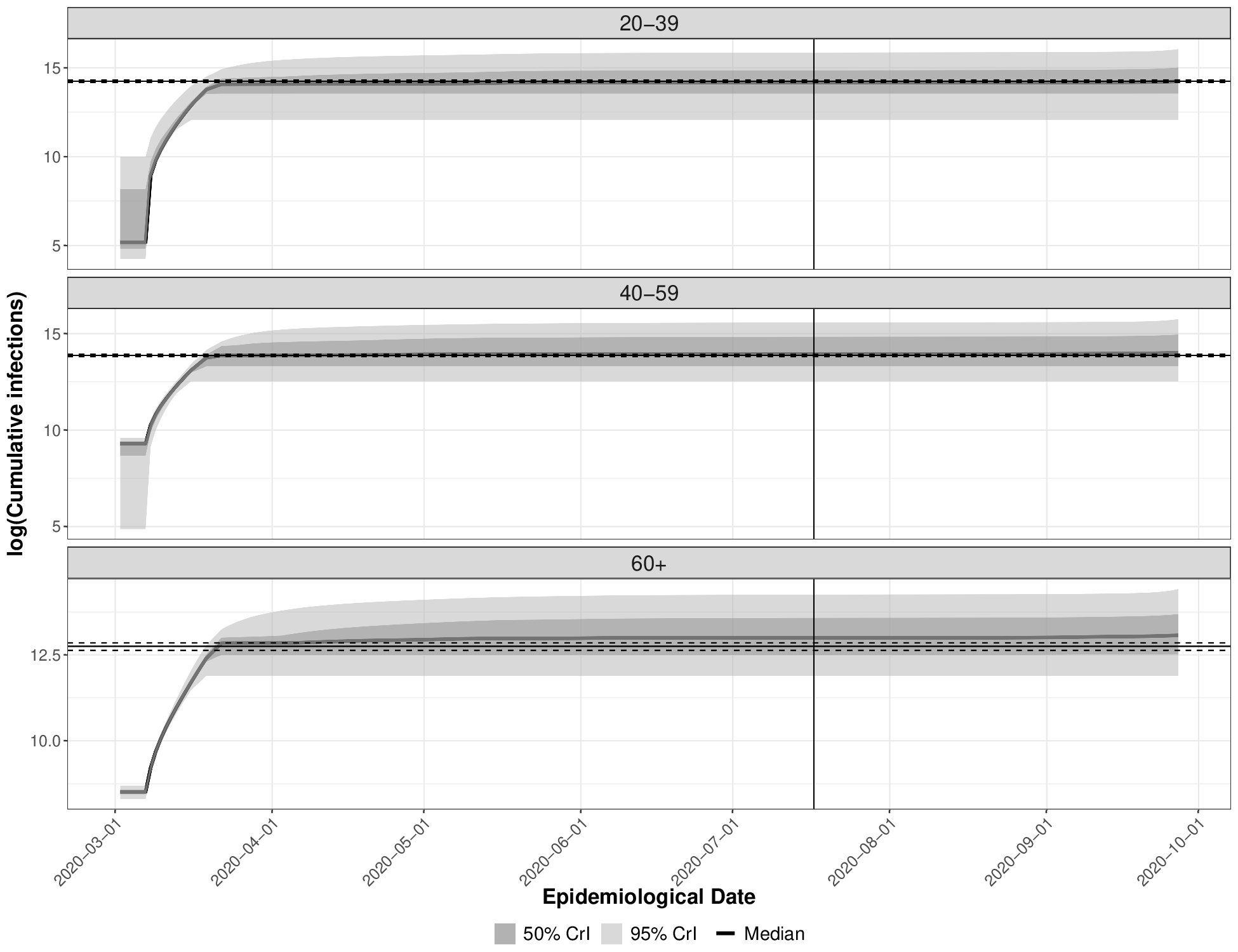}
\caption{COVID-19 epidemic in England - External validation. Log-transformed posterior median age-stratified cumulative infections (50\% and 95\% credible intervals, CrI) under the xBM model. 
The age-adjusted estimated counts for adults and the respective 95\% confidence intervals of cumulative infections based on REACT-2 \citep{ward_react2} are represented by horizontal lines. The vertical line corresponds to middle of July 2020.}
\label{fig:multitype_model_validation}
\end{figure}

Finally, we assess the xBM model by comparing its estimated age-stratified cumulative infections with estimates from the REACT-2 population survey \citep{ward_react2}.
Figure~\ref{fig:multitype_model_validation} shows good agreement between them for the adult age groups, demonstrating good agreement with an independent source of evidence.

\section{Discussion}\label{section:discussion}

In this paper, we introduce a hierarchical construction of exchangeable Gaussian processes that provides a parsimonious and interpretable specification within the broader class of coregionalisation and latent factor models. Motivated by the hierarchical structure of longitudinal data collected across related populations, the proposed formulation decomposes temporal variation into shared and population-specific components. This provides a natural mechanism for partial pooling, with intra-class correlation coefficients quantifying the degree of dependence across populations. The framework supports prediction for both observed and previously unobserved populations, accounting for the additional uncertainty associated with the latter. It also allows the degree of pooling to be learned from the data, ranging from a common latent process shared by all populations to independent population-specific processes.

The simulation study and empirical applications illustrate how the benefits of the proposed exchangeable GPs depend on the extent of dependence across populations. In the simulation study, the exchangeable formulations generally improve predictive performance when the underlying processes share temporal variation, even when the degree of cross-dependence is relatively weak. The three epidemic applications provide complementary evidence. For reproduction numbers across European countries, the exchangeable models improve predictive performance while revealing substantial heterogeneity in country-specific pooling parameters.  In the Chikungunya application, the evidence for shared transmission dynamics is weaker and differences in LOOIC among GP formulations are modest, but the ICCs still provide an interpretable assessment of cross-island dependence. In contrast, the age-stratified COVID-19 application reveals strong cross-dependence across age groups, accompanied by substantially better leave-one-out predictive performance for the exchangeable BM model relative to its independent counterpart. Together, these findings highlight the ability of the proposed framework to accommodate different degrees of cross-dependence while providing interpretable information about shared and population-specific temporal variation.

Beyond the applications considered here, the proposed generative framework could also support the investigation of alternative epidemic control strategies. In particular,
the nonlinear stochastic epidemic models considered in Section~6.3 provide a mechanism for simulating epidemic trajectories under different assumptions about transmission
dynamics. Combined with an appropriate model of intervention effects, such generative models could be incorporated into reinforcement learning approaches to epidemic control \citep{Giacomo}, allowing alternative intervention policies to be evaluated in terms of disease burden and socioeconomic costs. More generally, while our focus has been on epidemic modelling, the methodology extends naturally to a wide range of multilevel settings. These include longitudinal data analysis, where the proposed non-parametric Bayesian framework can accommodate unbalanced designs and missing observations, and multivariate time series in financial or econometric applications, where cross-sectional dependence, temporal autocorrelation, and covariate effects are of interest. A further direction concerns causal inference, including difference-in-differences and interrupted time series settings. In such applications, exchangeable GPs can be used to model untreated potential outcomes and construct counterfactual trajectories, borrowing information across related units through their shared latent structure. This provides a connection to factor-based approaches and recent developments in generalised Bayesian causal inference \citep{Angelos} and robust GP modelling \citep{FXrobGPs}.

\section*{Acknowledgments}
The authors are grateful to Michalis Titsias for helpful discussions on Gaussian Process Theory and Computation, and to the anonymous referees for their constructive and detailed comments that improved the content and presentation of the paper.

\section*{Funding}
Lampros Bouranis was supported by the European Union's Horizon 2020 research and innovation programme under the Marie Sklodowska-Curie grant agreement No 101027218.

 \noindent{\it Conflict of Interest}: None declared.

\section*{Supplementary Material}
The reader is referred to the online Supplementary Material for additional information regarding this work. The Supplementary Material consists of five sections. Section S1 provides Gaussian inference results for exchangeable processes, including marginal likelihoods, posterior prediction and inference for the shared latent process.
Section S2 presents Markov and state-space representations based on Brownian motion and Ornstein--Uhlenbeck processes, together with their computational implications. Section S3 provides implementation details, additional data information for the COVID-19 application, and Bayesian inference details, including prior specifications and MCMC
diagnostics. Section S4 presents complementary results from the simulation study, while Section S5 provides additional results for the real-data applications, including posterior predictive checks and estimated epidemic quantities.

\bibliographystyle{apalike}
\bibliography{xbm}


\end{document}